\documentclass[%
reprint,
 amsmath,amssymb,
 aps,
]{revtex4-2}

\usepackage{graphicx}
\usepackage{dcolumn}
\usepackage{bm}
\usepackage{hyperref}
\hypersetup{
    colorlinks=true, 
    linkcolor=blue, 
    citecolor=blue, 
    urlcolor=blue, 
}

\begin{document}


\title{Grassmann phase space dynamics of strongly-correlated fermions}

\author{Hassan Al-Hamzawi}
\author{Alessandro Principi}
\author{Leone Di Mauro Villari}
\affiliation{Department of Physics and Astronomy, University of Manchester, Manchester M13 9PL, UK}

\date{\today}

\begin{abstract}

We discuss the numerical implementation of two related representations of fermionic density matrices which have been introduced in \href{https://doi.org/10.1016/j.aop.2016.03.006}{Annals of Physics 370, 12 (2016)}.
In both of them, the density matrix is expanded in a basis of Bargmann coherent states with weights given by the two phase space distributions.
We derive the equations of motion for the distributions when imaginary time evolution is generated by the Hubbard Hamiltonian.
One of them is a Grassmann Fokker-Planck equation that can be re-cast into a remarkably simple It\^{o} form involving solely complex variables.
In spite of this simple form, we demonstrate that complications arise in numerically computing the expectation value of any observable.
These are due to exponential growth in the matrix elements of the stochastic propagator, delicate numerical sensitivity in performing primitive linear algebra operations, and the re-appearance of a sign problem.

\end{abstract}

\maketitle


\section{\label{sec:Introduction} Introduction}

The study of strongly-correlated systems is a notoriously difficult mathematical problem since interactions amongst all particles, and hence nonlinearities of the associated field theory, become fundamental in determining the state of the system. In this respect, practical applications favour descriptions that are amenable to numerical simulation. It is therefore pivotal to find new numerical methods that can overcome the limitations of existing techniques, such as the sign problem of quantum Monte Carlo calculations or the substantial overhead in iterant matrix product algorithms.
One way to achieve this is to represent the density matrix by expanding it in terms of an overcomplete basis, for example given by coherent states. In this way, states and operators take the form of functions and operations over them, respectively. Such operations preserve the operator algebra. 

The algebra of bosonic operators is based on commutators, therefore states and operators can be expressed as functions of (commuting) complex variables. This approach has been extensively studied and has led to numerous formulations, such as the Glauber and Sudarshan P distribution~\cite{Glauber1963, Sudarshan1963}, the Husimi Q distribution~\cite{Husimi1940}, and the Wigner W distribution~\cite{Wigner1932, Moyal1949}. These representations have been widely used to study strongly-interacting and heavily nonlinear bosonic systems in various fields, from quantum optics~\cite{Drummond:87} to Bose-Einstein condensation~\cite{Opanchuk_2012} and gravity analogues~\cite{Carusotto_2008}.

The algebra of fermionic operators on the other hand involves anti-commutators and cannot be represented in terms of complex variables. This complicates the computational simulation of fermionic systems since it is difficult to naturally absorb the operator algebra into the description. Calculations performed using complex variables must explicitly preserve exchange anti-symmetry. This produces an abundance of terms that have similar moduli and opposing signs and thus leads to a great deal of cancellation in their combined contributions ({\it i.e.} the infamous sign problem). A finer granularity and increased computational complexity are hence required to offset the loss in numerical precision.

Attempts have been made to adapt the techniques of coherent phase space distributions to fermions, as proposed by Cahill and Glauber~\cite{Cahill&Glauber1999}, and Dalton, Jeffers, and Barnett~\cite{PhaseSpaceMethods2014, Dalton&Jeffers&Barnett2016}. Analogous descriptions have been used to derive analytical results regarding two-level atoms interacting with electromagnetic fields~\cite{2LA2000, Qubit2005} as well as in the Jaynes-Cummings model~\cite{Dalton&Jeffers&Barnett2013}. Corney and Drummond have also shown that a different but related representation given in terms of Gaussian states has an equivalent formulation for fermions~\cite{Corney&Drummond2004, Corney&Drummond2005, Corney&Drummond2006}. 
The latter has been able to reproduce certain ground state properties of the Hubbard model~\cite{SymmetryProjection2005}, albeit with the aid of auxiliary symmetry projectors.

So far, Grassmann phase space formulations have had only very limited numerical implementations~\cite{Trapped1D} despite being the most natural and intuitive technique to represent fermionic quantum states as phase space distributions.
These techniques are also very promising since, supposedly, they do not suffer from some of the shortcomings of conventional methods, such as those mentioned above. For example, it can be shown that the statistical weights introduced by the Grassmann phase space distributions are related to the determinants of what are generally positive-definite matrices. This in turn implies that the entire formulation is in principle free of the infamous sign problem. Given these hopeful results, it is fundamental to test how far can these novel methods go, and to what extent they can enrich the present knowledge of the phase diagrams of strongly-correlated systems.

This article aims to extend the application of the Grassmann coherent state phase space developed in Dalton, Jeffers, and Barnett's work to the simulation of the Hubbard model, the ``gold standard'' of strongly-correlated physics.
In fact, the Hubbard model has been the centre of immense theoretical interest since its inception, due to no shortage of intriguing phenomenology, in particular its relation to high-temperature superconductivity. Whilst certain limiting cases have been extensively studied, the full phase diagram remains a contentious topic with many questions left unanswered~\cite{HubbardReview}. Due to the aforementioned intrigue, as well as its wide applicability in many-body quantum field theory, it serves as an ideal test-case to assess the suitability of any particular choice of representation.

In Sect.~\ref{sec:PhaseSpaceRepresentation}, we provide a brief description of fermionic coherent states, and a summary of several important results regarding the representation of physical density matrices as Grassmann phase space distributions.
In Sect.~\ref{sec:ThermalEvolution}, a pair of differential equations are derived for the imaginary time evolution of the grand-canonical ensemble. This is based on a choice of two available Grassmann representations, one of which results in a differential equation that has the Fokker-Planck structure.
In Sect.~\ref{sec:StochasticSampling}, the Fokker-Planck equation is expressed in a stochastic It\^{o} form using only complex numbers, and the extraction of quantum correlations from the statistical samples is addressed.
Finally, in Sect.~\ref{sec:AnalysisAndResults}, numerical results are presented in the case of a $2 \times 2$ square lattice. We show that unfortunately, despite the theoretical validity of the formulation, the stochastic dynamics results in the exponential growth in the propagator matrix elements. This introduces numerical instability into our implementation and reflects negatively on the accuracy and precision of observables.

\section{\label{sec:PhaseSpaceRepresentation} Phase Space Representation}

In this section we remind the reader of the nature of the basis used to parameterise the Hilbert space, {\it i.e.} the Bargmann coherent states, as well as of its properties.
We introduce fermion creation and annihilation operators $\hat{a}_i^\dagger$ and $\hat{a}_i$, respectively, where $i=1, \dots, n$ represents a set of quantum numbers. They satisfy the canonical anti-commutation relations. They also anti-commute with any Grassmann variable $g_i$. In formulae,
\begin{equation}
    \label{eq:FermionAnticommutators}
    \{ \hat{a}_i, \hat{a}_j^\dagger \} = \delta_{i, j} ~, \{ \hat{a}_i , \hat{a}_j \} = 0,
\end{equation}
and
\begin{equation}
\{ g_i, \hat{a}_j \} = \{ g_i, \hat{a}_j^\dagger \} = 0,
\end{equation}
where $\{ \cdot, \cdot \}$ is the anticommutator between pairs of operators or (Grassmann) variables.
The Grassmann variables commute with the vacuum state $g_i \left | 0 \right \rangle = \left | 0 \right \rangle g_i$. Thus, in the occupation basis Grassmann variables commute or anti-commute with even or odd occupancy states respectively. 
Hermitian conjugation is extended to also act on any Grassmann coefficients. It reverses the order of the symbols and acts to conjugate them
\begin{equation}
    \label{eq:CompositeConjugation}
    (g_i \hat{a}_j)^\dagger = \hat{a}_j^\dagger g_i^* .
\end{equation}
We emphasise that the conjugate of a Grassmann variable is itself an independent Grassmann variable, see App.~\ref{sec:GrassmannAlgebra} for more details.

\subsection{\label{subsec:BargmannCoherentStates} Bargmann Coherent States}

Bargmann coherent states~\cite{PhaseSpaceMethods2014, Dalton&Jeffers&Barnett2016} are defined by the action of the displacement operator $\hat{R}(\bm{g}) = \exp{(\sum_i \hat{a}_i^\dagger g_i)}$ on the vacuum state
\begin{equation}
    \label{eq:BargmannKet}
    \left | \bm{g} \right \rangle = \hat{R}(\bm{g}) \left | 0 \right \rangle.
\end{equation}
The bra is then defined as
\begin{equation}
 \label{eq:BargmannBra}
    \left \langle \bm{g} \right | = \left \langle 0 \right | \hat{R}(\bm{g})^\dagger ,
\end{equation}
where $\hat{R}(\bm{g})^\dagger = \exp{(\sum_i g_i^* \hat{a}_i)}$ and, as pointed out in the end of the previous section, $g_i$ and $g_i^*$ are independent variables.
By construction they commute with Grassmann variables. The action of creation and annihilation operators on these states is
\begin{gather}
    \label{eq:BargmannRelationsKet}
    \hat{a}_i \left | \bm{g} \right \rangle = g_i \left | \bm{g} \right \rangle ,~ \hat{a}_i^\dagger \left | \bm{g} \right \rangle = - \overrightarrow{\frac{\partial}{\partial g_i}} \left | \bm{g} \right \rangle = \left | \bm{g} \right \rangle \overleftarrow{\frac{\partial}{\partial g_i}} , \\
    \label{eq:BargmannRelationsBra}
    \left \langle \bm{g} \right | \hat{a}_i^\dagger = \left \langle \bm{g} \right | g_i^* ,~ \left \langle \bm{g} \right | \hat{a}_i = - \left \langle \bm{g} \right | \overleftarrow{\frac{\partial}{\partial g_i^*}} = \overrightarrow{\frac{\partial}{\partial g_i^*}} \left \langle \bm{g} \right | .
\end{gather}
Here the arrows make clear the direction in which the derivative operates (see also App.~\ref{subsec:GrassmannDifferentiation} for more detail about Grassmann differentiation).
Furthermore, as the conventional coherent states, they satisfy these relations
\begin{gather}
    \label{eq:BargmannInnerProduct}
    \left \langle \bm{u} | \bm{v} \right \rangle = \exp{(\bm{u}^* \cdot \bm{v})} , \\
    \label{eq:BargmannDyadTrace}
    \mathrm{Tr}(\left | \bm{v} \right \rangle \left \langle \bm{u} \right |) = \left \langle - \bm{u} | \bm{v} \right \rangle = \left \langle \bm{u} | - \bm{v} \right \rangle , \\
    \label{eq:IdentityResolution}
    \hat{I} = \int dg_n^* dg_n \dots dg_1^* dg_1 \ \exp{(\bm{g} \cdot \bm{g}^*)} \left | \bm{g} \right \rangle \left \langle \bm{g} \right | ,
\end{gather}
which in turn imply that the Bargmann states form an overcomplete basis for the Hilbert space. Note that, since $g_i$ and $g_i^*$ are independent variables for all $i=1, \dots, n$, the integration in Eq.~(\ref{eq:IdentityResolution}) runs over both sets of Grassmann variables.

\subsection{\label{subsec:GrassmannDistributionFunctions} Grassmann Distribution Functions}

Dalton, Jeffers, and Barnett~\cite{PhaseSpaceMethods2014, Dalton&Jeffers&Barnett2016} demonstrate that any physical fermionic density matrix can be uniquely expanded as a distribution over Bargmann coherent states. Using Eq.~(\ref{eq:IdentityResolution}), the density matrix becomes
\begin{equation}
    \label{eq:DistributionIntegral}
    \hat{\rho} = \int d^n\bm{u} \ d^n\bm{v}^* B(\bm{u}, \bm{v}^*) \left | \bm{u} \right \rangle \left \langle \bm{v} \right | ,
\end{equation}
where $\bm{u}$ and $\bm{v}$ are two unrelated sets of Grassmann variables (see also App.~\ref{subsec:GrassmannIntegration} for more detail about Grassmann integration) and
\begin{equation}
    \label{eq:DistributionIntegral2}
    B(\bm{u}, \bm{v}^*) = \int d^n\bm{u}^* d^n\bm{v} \left \langle \bm{u} | \hat{\rho} | \bm{v} \right \rangle e^{\bm{u} \cdot \bm{u}^*} e^{\bm{v} \cdot \bm{v}^*}.
\end{equation}
Defining $\bm{g} = \bm{u}$ and $\bm{\tilde{g}} = \bm{v}^*$, as well as the pair of normalised and un-normalised projector operators
\begin{equation}
    \label{eq:BargmannProjectors}
    \hat{\Lambda}^{(n)}(\bm{g}, \bm{\tilde{g}} ) = \frac{\left | \bm{g} \right \rangle \left \langle \bm{\tilde{g}}^* \right |}{\mathrm{Tr}(\left | \bm{g} \right \rangle \left \langle \bm{\tilde{g}}^* \right |)} , \ \hat{\Lambda}(\bm{g}, \bm{\tilde{g}} ) = \left | \bm{g} \right \rangle \left \langle \bm{\tilde{g}}^* \right | ,
\end{equation}
it is possible to re-express the density matrix in two equivalent forms
\begin{align}
    \label{eq:DensityUnNormalised}
    \hat{\rho} & = \int d^n\bm{g} \ d^n\bm{\tilde{g}} \ B(\bm{g}, \bm{\tilde{g}}) \hat{\Lambda}(\bm{g}, \bm{\tilde{g}}) \\
    \label{eq:DensityNormalised}
    & = \int d^n\bm{g} \ d^n\bm{\tilde{g}} \ P(\bm{g}, \bm{\tilde{g}}) \hat{\Lambda}^{(n)}(\bm{g}, \bm{\tilde{g}}) .
\end{align}
Both distribution functions are Grassmann even, with the $B$-distribution expressed as the Grassmann integral in Eq.~(\ref{eq:DistributionIntegral2}) and related to $P$-distribution through
\begin{equation}
    \label{eq:NormalisedUnNormalised}
    P(\bm{g}, \bm{\tilde{g}}) = B(\bm{g}, \bm{\tilde{g}}) e^{\bm{g} \cdot \bm{\tilde{g}}} .
\end{equation}

The expectation value of any normally ordered product of creation and annihilation operators can be evaluated through a Grassmann integral weighted by the $P$-distribution function as
\begin{eqnarray}
    \begin{aligned}
    \label{eq:Trace}
    \mathrm{Tr} [\hat{a}_{\mu_1}^\dagger \dots \hat{a}_{\mu_p}^\dagger \hat{a}_{\nu_q} \dots \hat{a}_{\nu_1} \hat{\rho}]
    &=\int d^n\bm{g} \ d^n\bm{\tilde{g}} \ P(\bm{g}, \bm{\tilde{g}}) \\
    &\times
    g_{\nu_q} \dots g_{\nu_1} 
    \tilde{g}_{\mu_1} \dots \tilde{g}_{\mu_p} .
    \end{aligned}
\end{eqnarray}

Finally, via the integral expansion of the density matrix given in Eqs.~(\ref{eq:DensityUnNormalised}) and~(\ref{eq:DensityNormalised}), the relations stated in Eqs.~(\ref{eq:BargmannRelationsKet}) and~(\ref{eq:BargmannRelationsBra}), and the integration by parts formulae found in Eqs.~(\ref{eq:LeftByParts}) and~(\ref{eq:RightByParts}), it is possible to show the following operator to phase space mappings
\begin{eqnarray}
\begin{gathered}
    \label{eq:P_correspondences}
    \hat{a}_i \hat{\rho} \to g_i P , \\
    \hat{\rho} \hat{a}_i \to P \left (\overleftarrow{\frac{\partial}{\partial \tilde{g}_i}} - g_i \right ) , \\
    \hat{a}_i^\dagger \hat{\rho} \to \left (\overrightarrow{\frac{\partial}{\partial g_i}} - \tilde{g}_i \right ) P , \\
    \hat{\rho} \hat{a}_i^\dagger  \to P \tilde{g}_i ,
\end{gathered}
\end{eqnarray}
for the $P$-distribution
and
\begin{eqnarray}
\begin{gathered}
    \label{eq:B_correspondences}
    \hat{a}_i \hat{\rho} \to  g_i B , \\
    \hat{\rho} \hat{a}_i \to  B \overleftarrow{\frac{\partial}{\partial \tilde{g}_i}} , \\
    \hat{a}_i^\dagger \hat{\rho} \to \overrightarrow{\frac{\partial}{\partial g_i}} B , \\
    \hat{\rho} \hat{a}_i^\dagger  \to B \tilde{g}_i ,
\end{gathered}
\end{eqnarray}
for the $B$-distribution.
In the case of sequences of several operators, the previous correspondences are applied consecutively in order of proximity to the density matrix.

\section{\label{sec:ThermalEvolution} Imaginary Time Evolution}

The Hubbard model~\cite{Hubbard1963, Gutzwiller1963, Kanamori1963} is one of the most established and well studied formulations of interacting fermions on a lattice. It is described by the Hamiltonian
\begin{equation}
    \label{eq:HubbardHamiltonian}
    \hat{H} = - \sum\limits_{i,j, \sigma} \bm{\tau}_{i, j} \hat{a}_{i, \sigma}^\dagger \hat{a}_{j, \sigma} + U \sum\limits_k \hat{a}_{k, \uparrow}^\dagger \hat{a}_{k, \uparrow} \hat{a}_{k, \downarrow}^\dagger \hat{a}_{k, \downarrow} ,
\end{equation}
where $\bm{\tau}$ is the hopping matrix. Its elements represent hopping amongst nearest neighbours, next-to-nearest neighbours, and so on. In this paper we will consider only the former. We parametrise the hopping between any pair of neighbouring sites $(i, j)$ with the amplitude $t$.
In Eq.~(\ref{eq:HubbardHamiltonian}), $\sigma$ denotes the particle spin and takes on the symbolic values $\uparrow \downarrow$, while $U$ is the on-site interaction potential. For later factorisation convenience, it is possible to re-cast the second term of the Hamiltonian in the form of a normally ordered product as
\begin{equation}
    \label{eq:InteractionHamiltonian}
    \hat{H}_I = \sum\limits_{i, j, \sigma, \sigma '} \bm{U}_{i \sigma, j \sigma '} \hat{a}_{i, \sigma}^\dagger \hat{a}_{j, \sigma '}^\dagger \hat{a}_{j, \sigma '} \hat{a}_{i, \sigma} ,
\end{equation}
where
\begin{equation}
    \label{eq:InteractionMatrix}
    \bm{U}_{i \sigma, j \sigma '} = - \frac{| U |}{2} \sum\limits_k \delta_{i, k} \kappa(\sigma) \delta_{j, k} \kappa(\sigma ') ,
\end{equation}
and
\begin{equation}
    \kappa(\sigma) =
    \begin{cases}
        1 & \sigma = \uparrow \\
        - \mathrm{sign}(U) & \sigma = \downarrow
    \end{cases} .
\end{equation}

To derive the equations of motion satisfied by the $P$- and $B$-distributions, we note that the density matrix for the grand-canonical ensemble satisfies the following Matsubara differential equation
\begin{equation}
    \label{eq:GrandCanonicalEnsemble}
    \frac{d}{d\beta} \hat{\rho} = - \frac{1}{2} \{ \hat{H} - \mu \hat{N}, \hat{\rho} \} ,
\end{equation}
where $\beta^{-1} = k_B T$. Henceforth, we absorb the chemical potential $\mu$ into the diagonal elements of the hopping matrix $\bm{\tau}$ to simplify our notation.

Appendix~\ref{sec:DifferentialEquations} shows how the correspondences of Eqs.~(\ref{eq:P_correspondences}) can be utilised to derive the differential equation describing the dynamics of the $P$-distribution. The key result is given in Eq.~(\ref{eq:ThermalNormalised}), which represents a family of coupled differential equations relating the complex valued coefficients of the multinomial associated with $P$. The presence of terms with a mismatched number of Grassmann multiplications and differentiations hinders the ability to re-express the differential equation in any form amenable to numerical integration.

The $B$-distribution on the other hand has simpler correspondences, as shown in Eqs.~(\ref{eq:B_correspondences}). Using them, it is possible to show that it satisfies a relation that has the general structure of a Fokker-Planck equation, though the coefficients are Grassmann variables or products thereof [see  Eq.~(\ref{eq:ThermalUnNormalised}) derived in App.~\ref{sec:DifferentialEquations}]. With summation implied over all the present indices, the differential equation is re-written in a compact form as
\begin{eqnarray}
\begin{gathered}
    \label{eq:FokkerPlanck}
    \frac{d}{d\beta} B = - \left (B A_{i \sigma}^\mu \right ) \overleftarrow{\frac{\partial}{\partial g_{i \sigma}^\mu}} + \frac{1}{2} \left (B D_{i \sigma, j \sigma '}^{\mu, \nu} \right ) \overleftarrow{\frac{\partial}{\partial g_{j \sigma '}^\nu}} \overleftarrow{\frac{\partial}{\partial g_{i \sigma}^\mu}} , \\
    \label{eq:FokkerPlanckDrift}
    A_{i \sigma}^\mu = - \frac{1}{2} \sum\limits_{j, \sigma ', \nu} \delta_{\sigma, \sigma '}^{\mu, \nu} \left (\delta^{\nu, 1} \bm{\tau}_{i, j} + \delta^{\nu, 2} \bm{\tau}_{i, j}^* \right ) g_{j \sigma '}^\nu , \\
    \label{eq:FokkerPlanckDiffusion}
    D_{i \sigma, j \sigma '}^{\mu, \nu} = \frac{| U |}{2} \delta^{\mu, \nu} \sum\limits_k \delta_{i, k} g_{i \sigma}^\mu \kappa(\sigma) \delta_{j, k} g_{j \sigma '}^\nu \kappa(\sigma ') ,
\end{gathered}
\end{eqnarray}
where the notation $g^\mu$ has been introduced to distinguish the two sets of Grassmann variables, {\it i.e.} $g^1 = \bm{g}$ and $g^2 = \bm{\tilde{g}}$. Despite removing the mismatch in the number of Grassmann multiplications and differentiations, Eq.~(\ref{eq:FokkerPlanck}) viewed directly still represents a family of coupled differential equations. However, this current form presents the opportunity for a stochastic transformation.

Considering any arbitrary Grassmann functions $F(\bm{g}, \bm{\tilde{g}})$ and $H(\bm{g}, \bm{\tilde{g}}) = F(\bm{g}, \bm{\tilde{g}}) e^{\bm{g} \cdot \bm{\tilde{g}}}$ then their phase space distribution expectations are related through
\begin{eqnarray}
\begin{aligned}
    \label{eq:Expectation}
    \overline{F(\bm{g}, \bm{\tilde{g}})}|_P & = \int d^n\bm{g} \ d^n\bm{\tilde{g}} \ F(\bm{g}, \bm{\tilde{g}}) P(\bm{g}, \bm{\tilde{g}}, \beta) \\
    & = \int d^n\bm{g} \ d^n\bm{\tilde{g}} \ F(\bm{g}, \bm{\tilde{g}}) B(\bm{g}, \bm{\tilde{g}}, \beta) e^{\bm{g} \cdot \bm{\tilde{g}}} \\
    & = \overline{H(\bm{g}, \bm{\tilde{g}})}|_B .
\end{aligned}
\end{eqnarray}
Taking the derivative of Eq.~(\ref{eq:Expectation}) with respect to $\beta$ (which only acts on the distribution function), and using the Fokker-Planck relation in Eq.~(\ref{eq:FokkerPlanck}) we obtain
\begin{eqnarray}
    \begin{aligned}
    \label{eq:ExpectationDerivative}
    \frac{d}{d\beta} \overline{H(\bm{g}, \bm{\tilde{g}})}|_B & = \int d^n\bm{g} \ d^n\bm{\tilde{g}} \ H(\bm{g}, \bm{\tilde{g}}) \frac{d}{d\beta}B(\bm{g}, \bm{\tilde{g}}, \beta) \\
    & = \int d^n\bm{g} \ d^n\bm{\tilde{g}} \ [- \sum\limits_{i, \sigma, \mu} H \overleftarrow{\frac{\partial}{\partial g_{i \sigma}^\mu}} A_{i \sigma}^\mu \\
    & + \frac{1}{2} \sum\limits_{i, j, \sigma, \sigma ', \mu, \nu} H \overleftarrow{\frac{\partial}{\partial g_{j \sigma '}^\nu}} \overleftarrow{\frac{\partial}{\partial g_{i \sigma}^\mu}} D_{i \sigma, j \sigma '}^{\mu, \nu} ] B .
    \end{aligned}
\end{eqnarray}
Here, we integrated by parts using the standard rules of Grassmann calculus, which are provided in Eqs.~(\ref{eq:RightDerivativeAntiCommutativity}) and~(\ref{eq:RightByParts}) for convenience. Eq.~(\ref{eq:ExpectationDerivative}) is a useful relation to derive stochastic evolution in the following section.

\section{\label{sec:StochasticSampling} Stochastic Sampling}

Instead of considering the expectation value of a fixed Grassmann function $H(\bm{g}, \bm{\tilde{g}})$ and an evolving distribution $B(\bm{g}, \bm{\tilde{g}}, \beta)$, stochastic calculus is employed to facilitate sampling. The Grassmann variables themselves are allowed to evolve under a static distribution $B(\bm{g}, \bm{\tilde{g}})$ which does not depend explicitly on $\beta$ (though it can depend on it through $\bm{g}$ and $\bm{\tilde{g}}$). The evolution of such variables, which we denote as $g_{i \sigma}^\mu (\beta)$, is such that it ensures that Eq.~(\ref{eq:ExpectationDerivative}) continues to hold in the mean.

Expanding the change of Eq.~(\ref{eq:Expectation}) to second order using the rules of Grassmann calculus recalled in Eq.~(\ref{eq:RightTaylor}) for convenience, we obtain
\begin{eqnarray}
    \begin{aligned}
    \label{eq:ExpectationDifference}
    \overline{\delta H} & = \int d^n\bm{g} \ d^n\bm{\tilde{g}} \ (H (\beta + \delta \beta) - H (\beta)) B(\bm{g}, \bm{\tilde{g}}) \\
    &= \int d^n\bm{g} \ d^n\bm{\tilde{g}} \ [\sum\limits_{i, \sigma, \mu} H \overleftarrow{\frac{\partial}{\partial g_{i \sigma}^\mu}} \delta g_{i \sigma}^\mu \\
    & + \frac{1}{2} \sum\limits_{i, j, \sigma, \sigma ', \mu, \nu} H \overleftarrow{\frac{\partial}{\partial g_{j \sigma '}^\nu}} \overleftarrow{\frac{\partial}{\partial g_{i \sigma}^\mu}} \delta g_{i \sigma}^\mu \delta g_{j \sigma '}^\nu ] B(\bm{g}, \bm{\tilde{g}}) .
    \end{aligned}
\end{eqnarray}
We assume a Langevin form for the stochastic differential equation satisfied by the Grassmann variables, {\it i.e.}
\begin{equation}
    \label{eq:GrassmannStochastic}
    \delta g_{i \sigma}^\mu = \mathcal{A}_{i \sigma}^\mu \delta \beta + \sum\limits_k \mathcal{B}_{i \sigma}^{\mu, k} \delta W_k + \mathcal{O}(\delta \beta^2),
\end{equation}
where $\delta W_k$ are independent Wiener increments~\cite{StochasticHandbook}, {\it i.e.} they are normally distributed random variables with zero mean and variance $\delta \beta$. Thus the averages of one and two Grassmann variable variations satisfy
\begin{eqnarray}
\begin{gathered}
    \label{eq:StochasticAverages}
    \left \langle \delta g_{i \sigma}^\mu \right \rangle_\text{Stochastic} = \mathcal{A}_{i \sigma}^\mu \delta \beta + \mathcal{O}(\delta \beta^2) , \\
    \left \langle \delta g_{i \sigma}^\mu \delta g_{j \sigma '}^\nu \right \rangle_\text{Stochastic} = \sum\limits_k \mathcal{B}_{i \sigma}^{\mu, k} \mathcal{B}_{j \sigma '}^{\nu, k} \delta \beta + \mathcal{O}(\delta \beta^2) ,
\end{gathered}
\end{eqnarray}
where $\left \langle \dots \right \rangle_\text{Stochastic}$ is the average over noise realisations.
Then, assuming that 
\begin{equation}
    \label{eq:StochasticFokkerPlanckEquivalences}
    A_{i \sigma}^\mu = -\mathcal{A}_{i \sigma}^\mu , \ D_{i \sigma, j \sigma '}^{\mu, \nu} = \sum\limits_k \mathcal{B}_{i \sigma}^{\mu, k} \mathcal{B}_{j \sigma '}^{\nu, k} ,
\end{equation}
it ensures that Eq.~(\ref{eq:ExpectationDifference}) reduces to Eq.~(\ref{eq:ExpectationDerivative}) in the mean and to leading order in $\delta \beta$.

Combining Eq.~(\ref{eq:FokkerPlanck}) with Eqs.~(\ref{eq:GrassmannStochastic}) and~(\ref{eq:StochasticFokkerPlanckEquivalences}), the infinitesimal evolution is finally expressed in terms of the following linear operators
\begin{eqnarray}
\begin{gathered}
    \label{eq:StochasticLinearOperators}
    \bm{g}_\uparrow (\beta + \delta \beta) = \left(I + \frac{\delta \beta}{2} \bm{\tau} + \sqrt{\frac{| U |}{2}} R \right) \bm{g}_\uparrow (\beta) , \\
    \bm{g}_\downarrow (\beta + \delta \beta) = \left(I + \frac{\delta \beta}{2} \bm{\tau} - \mathrm{sign}(U) \sqrt{\frac{| U |}{2}} R \right) \bm{g}_\downarrow (\beta) , \\
    \bm{\tilde{g}}_\uparrow (\beta + \delta \beta) = \left(I + \frac{\delta \beta}{2} \bm{\tau}^* + \sqrt{\frac{| U |}{2}} \tilde{R} \right) \bm{\tilde{g}}_\uparrow (\beta) , \\
    \bm{\tilde{g}}_\downarrow (\beta + \delta \beta) = \left(I + \frac{\delta \beta}{2} \bm{\tau}^* - \mathrm{sign}(U) \sqrt{\frac{| U |}{2}} \tilde{R} \right) \bm{\tilde{g}}_\downarrow (\beta) ,
\end{gathered}
\end{eqnarray}
where $I$ is the identity matrix, $\bm{\tau}$ is the hopping matrix (which also includes the chemical potential), and $R$ and $\tilde{R}$ are both diagonal matrices whose entries are independent Wiener increments.
Integrating Eq.~(\ref{eq:StochasticLinearOperators}) over the finite interval between $\beta_0$ and $\beta$, we obtain
\begin{equation}
    \label{eq:Propagator}
    \bm{g}^\mu (\beta) = U^\mu (\beta ; \beta_0) \bm{g}^\mu (\beta_0) ,
\end{equation}
where $U^\mu (\beta ; \beta_0)$ is the imaginary time- ($\beta$-)ordered product of the operators above.

Having established a relation for the dynamics of the Grassmann variables, the static phase space distribution remains to be determined. Choosing initial conditions at $\beta_0 = 0$ such that $\hat{\rho} \propto \hat{I}$ and evaluating the Grassmann integral in Eq.~(\ref{eq:DistributionIntegral}), the $B$-distribution acquires the form
\begin{equation}
    \label{eq:InitialUnNormalised}
    B(\bm{g}_0, \bm{\tilde{g}}_0) \propto \exp{(\bm{g}_0 \cdot \bm{\tilde{g}}_0)} .
\end{equation}
where $\bm{g}_0 = (\bm{g}_{0 \uparrow}^T, \bm{g}_{0 \downarrow}^T)^T$ and $\bm{\tilde{g}}_0 = (\bm{\tilde{g}}_{0 \uparrow}^T, \bm{\tilde{g}}_{0 \downarrow}^T)^T$.

\begin{figure*}
    \centering
    \includegraphics[width=\linewidth, keepaspectratio]{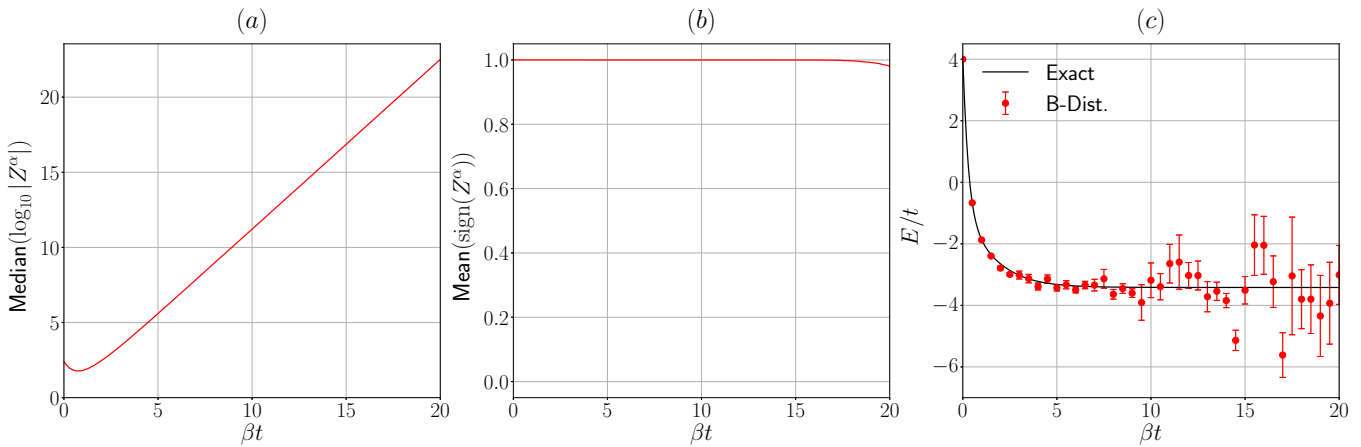}
    \caption{Ensemble statistics of $10^5$ samples on a $2 \times 2$ lattice with $U = 4t$, $\mu = 0$. The stochastic process realisations were generated via Euler-Maruyama integration at a step size of $\delta \beta t= 10^{-4}$. (a) Exponential growth of the modulus of the statistical weights. (b) Fluctuations in the sign of the statistical weights. (c) Expectation value of the Hamiltonian as extrapolated from the ensemble correlations.}
    \label{fig:GrassmannMu000}
\end{figure*}

We can now introduce the characteristic function to obtain the normally ordered correlations. Utilising two further sets of Grassmann variables $\bm{u}$ and $\bm{v}$, the characteristic function is defined to be
\begin{equation}
    \label{eq:CharacteristicTrace}
    \chi(\bm{u}, \bm{v}) = \mathrm{Tr}\left[\exp{\left(i \sum\limits_n v_n \hat{a}_n\right)} \hat{\rho} \exp{\left(i \sum\limits_m \hat{a}_m^\dagger u_m\right)} \right] .
\end{equation}
Recalling Eqs.~(\ref{eq:NormalisedUnNormalised}) and~(\ref{eq:Trace}), Eq.~(\ref{eq:CharacteristicTrace}) reduces to 
\begin{eqnarray} \label{eq:CharacteristicTrace2}
    \begin{aligned} 
    \chi(\bm{u}, \bm{v}) &= \int d^n\bm{g} \ d^n\bm{\tilde{g}} \ B(\bm{g}, \bm{\tilde{g}}) e^{(\bm{g} \cdot \bm{\tilde{g}} + i \bm{v} \cdot \bm{g} + i \bm{\tilde{g}} \cdot \bm{u})} \\
    &= \overline{\exp{(\bm{g} \cdot \bm{\tilde{g}} + i \bm{v} \cdot \bm{g} + i \bm{\tilde{g}} \cdot \bm{u})}}|_B .
    \end{aligned}
\end{eqnarray}
Hence, correlations can be obtained as derivatives of the characteristic function. The Grassmann variables in Eq.~(\ref{eq:CharacteristicTrace2}) are evaluated at imaginary time $\beta$. Using Eq.~(\ref{eq:Propagator}), we can rewrite Eq.~(\ref{eq:CharacteristicTrace2}) in terms of Grassmann variables at imaginary time $\beta_0$ as 
\begin{eqnarray}
\begin{gathered}
    \chi (\bm{u}, \bm{v}) = \overline{\exp{(\bm{g} \cdot \bm{\tilde{g}} + i \bm{v} \cdot \bm{g} + i \bm{\tilde{g}} \cdot \bm{u})}}|_\beta \\
    =
    \int d^n\bm{g}_0 \ d^n\bm{g}_0 \ e^{({\bm{g}}_0^T U^T \tilde{U} \bm{\tilde{g}}_0 + i \bm{v}^T U \bm{g}_0 + i {\bm{\tilde{g}}}_0^T \tilde{U}^T \bm{u})} B(\bm{g}_0, \bm{\tilde{g}}_0) ,
\end{gathered}
\end{eqnarray}
with $B(\bm{g}_0, \bm{\tilde{g}}_0)$ being the exponential in Eq.~(\ref{eq:InitialUnNormalised}). This has the form of a Gaussian-type integral over Grassmann variables and can be shown to give~\cite{QFT2002}
\begin{equation}
    \label{eq:CharacteristicEvaluated}
    \chi(\bm{u}, \bm{v}) \propto \mathrm{det}(I + U^T \tilde{U}) e^{(\bm{u}^T \tilde{U} (I + U^T \tilde{U})^{-1} U^T \bm{v})} .
\end{equation}

Integration of Eq.~(\ref{eq:Propagator}) produces a statistical ensemble of trajectories, each associated with its own set of moments. The dynamics, chosen to preserve Eq.~(\ref{eq:ExpectationDerivative}), ensures that the ensemble average converges to the true physical correlation.

Let the realisations of the stochastic process be indexed by $\alpha$. Identifying $Z^\alpha = \mathrm{det}(I + U^T_\alpha \tilde{U}_\alpha)$ and $M^\alpha = \tilde{U}_\alpha (I + U^T_\alpha \tilde{U}_\alpha)^{-1} U^T_\alpha$, where $U_\alpha$ and $\tilde{U}_\alpha$ are the evolution operators for a given realisation, the first order correlations can be expressed as
\begin{equation}
    \label{eq:FirstOrderCorrelations}
    \langle \hat{a}_{i, \sigma}^\dagger \hat{a}_{j, \sigma '} \rangle = \frac{\mathrm{Tr}[\hat{a}_{i, \sigma}^\dagger \hat{a}_{j, \sigma '} \hat{\rho}]}{\mathrm{Tr}[\hat{\rho}]} = \frac{\sum_\alpha Z^\alpha M^\alpha_{i \sigma , j \sigma '}}{\sum_\alpha Z^\alpha} ,
\end{equation}
with similar relations for the higher order moments obtained by expanding Eq.~(\ref{eq:CharacteristicTrace}) and Eq.~(\ref{eq:CharacteristicEvaluated}). Since Wiener increments at different imaginary times are uncorrelated, the average $\langle I + U^T_\alpha \tilde{U}_\alpha \rangle_\text{Stochastic}$ is a positive-definite matrix. This suggests that, excluding outliers, the statistical weights appearing above ought to generally remain positive throughout the ensemble.

\section{\label{sec:AnalysisAndResults} Analysis and Results}
\begin{figure*}
    \centering
    \includegraphics[width=\linewidth, keepaspectratio]{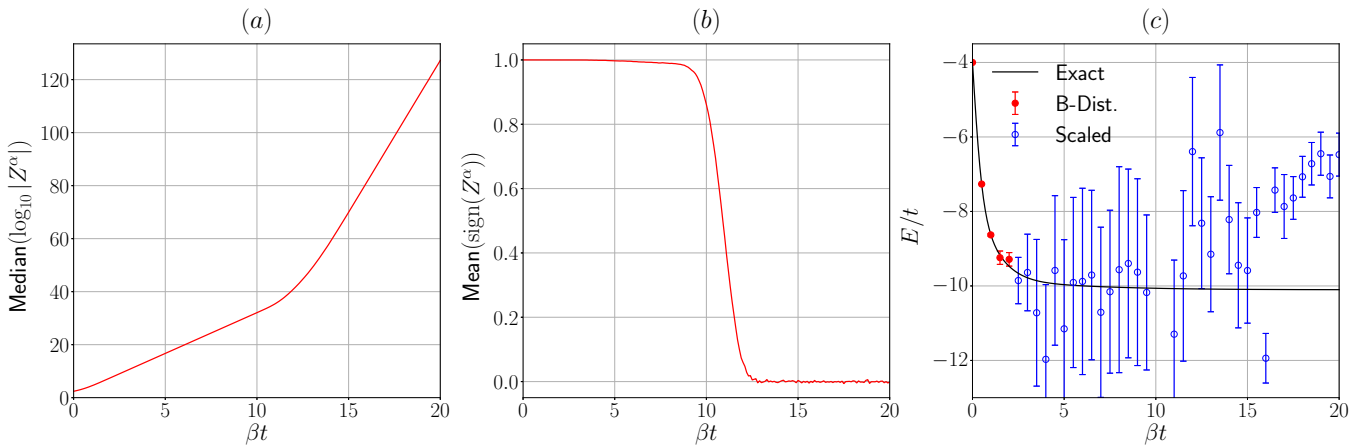}
    \caption{Ensemble statistics of $10^5$ samples on a $2 \times 2$ lattice with $U = 4t$, $\mu = 2t$. The stochastic process realisations were generated via Euler-Maruyama integration at a step size of $\delta \beta t= 10^{-4}$. To enhance readability, the (blue) scaled bars for $\beta t \geq 2.5$ represents half the base-10 logarithm of the calculated error.}
    \label{fig:GrassmannMu050}
\end{figure*}

With all the necessary frameworks established, we now apply the formalism to the case of a $2 \times 2$ square lattice. This system is of a sufficiently small size that the stochastic sampling may be contrasted with the results of an exact diagonalisation.

We first discuss the results obtained in the symmetric configuration with $\mu = 0$. Before addressing the behaviour of physical quantities, we highlight that Eq.~(\ref{eq:StochasticLinearOperators}) implies that the propagator matrix elements grow asymptotically in an exponential manner. This is illustrated in Fig.~\ref{fig:GrassmannMu000}(a), where we plot the median statistical weight of the trajectories $Z^\alpha$ as a function of $\beta t$ ($t$ being the hopping amplitude). This highlights that the reliability of the results could be particularly dependent on the precision of the underlying computer arithmetic system. Thus, the evaluation of determinants as well as the inversion of matrices can become numerically unstable operations.
As a consequence, $Z^\alpha$ cannot be guaranteed to have a fixed sign for all values of $\beta t$. We plot the average sign of $Z^\alpha$ in Fig.~\ref{fig:GrassmannMu000}(b), which is found to remain positive for all values of $\beta t$, though fluctuations do begin to arise briefly before the simulation cutoff.
Fig.~\ref{fig:GrassmannMu000}(c) showcases the expectation value of the Hamiltonian. Some deviations from the exact treatment are observed in addition to a steadily increasing uncertainty as $\beta$ increases.

The quality of the results drastically change when applied at half-filling ($\mu = 2t$). The rate of exponential growth demonstrated in Fig.~\ref{fig:GrassmannMu050}(a) is substantially steeper than the previous configuration. In addition, we find that $Z^\alpha$ has equal probability to be either positive or negative beyond a certain value of $\beta t$. This is shown in Fig.~\ref{fig:GrassmannMu050}(b), where the mean sign of $Z^\alpha$ is zero above $\beta t\gtrsim 12$. While for small values of $\beta t$ the statistical weight is positive for all trajectories, its sign gets rapidly randomised as $\beta t$ increases. Thus a sign problem re-emerges.
Both the aforementioned issues seem to be inherent attributes of this description of the system. Employing more sophisticated integration techniques and increasing the lattice size not only failed to alleviate the instability but rather made it more pronounced. 
We note that while $Z^\alpha$ cannot be guaranteed to remain of fixed sign indefinitely, the fact that $\langle I + U^T_\alpha \tilde{U}_\alpha \rangle_\text{Stochastic}$ is positive definite suggests that the sign problem might itself be a symptom of numerical instability in the evaluation of the determinant rather than an independent issue entirely.

In order to systematically account for these shortcomings and their influence over the accuracy and precision of the calculations, the correlations are re-defined to be
\begin{equation}
    \label{eq:ModifiedFirstOrderCorrelations}
    \langle \hat{a}_{i, \sigma}^\dagger \hat{a}_{j, \sigma '} \rangle = \left. \frac{\sum_\alpha Z^\alpha M^\alpha_{i \sigma , j \sigma '}}{\sum_\alpha | Z^\alpha |} \middle/ \frac{\sum_\alpha Z^\alpha}{\sum_\alpha | Z^\alpha |} \right. ,
\end{equation}
where both the numerator and the denominator are to be evaluated separately.

This modified expectation value of the Hamiltonian is shown in Fig.~\ref{fig:GrassmannMu050}(c). There, we find that the average energy starts fluctuating at a relatively small value of $\beta t$ due to the steady loss of precision in matrix inversion combined with the strongly correlated dynamics. Furthermore, the reliability of the results rapidly deteriorates as shown by the larger and larger error bars and as expected from the prior observations. We thus conclude that, for the most interesting cases, the Grassmann phase space representation of the Hubbard model, though mathematically valid, falls short of providing the level of numerical stability desirable for practical  applications.

\section{\label{sec:Conclusions} Conclusions}

The theory of Grassmann phase space distributions was introduced and applied to the grand-canonical ensemble of the Hubbard model. Two differential equations were obtained to describe the dynamics of two related phase space distributions in imaginary time. The $B$-distribution produced a Fokker-Planck differential equation which can be re-cast into a stochastic evolution of Grassmann variables.

Several critical limitations were found in performing numerical calculations in this formulation. Firstly, the matrix elements of the stochastic propagator were shown to grow exponentially. Secondly, the statistical weight associated to each trajectory exhibited severe fluctuations in sign that produced catastrophic cancellation. The uncertainty these findings introduced to ensemble correlations was most critical in the non-perturbative parameter regimes, where other numerical methods also face similar convergence issues. Hence, our results show that any extension of the phase diagram beyond what is presently known is not possible, at least within our current implementation.

Despite the numerical fragility encountered in this investigation, the Grassmann phase space theory still demonstrates the possibility of mapping Hamiltonians containing two-body interactions into stochastic differential equations that are fitting for computational simulation. Its application is deemed most suitable in circumstances involving a small number of interacting modes. These are cases typically encountered in the quantum optics of two-level systems, when the initial distribution is considerably restrictive in its non-zero correlations, thereby sidestepping most of the highlighted issues, or when the dynamics of the $P$-distribution renders it directly amenable to linear stochastic methods.

Akin to similar probabilistic techniques, this approach achieves its best accuracy at high temperatures (small imaginary times) before the development of numerical instabilities. It is however necessary to ensure that the spectra of the operators in Eq.~(\ref{eq:StochasticLinearOperators}) remains bounded when taking the thermodynamic limit. Otherwise, the exponential run-off would present a problem even in such a semi-classical regime.

\appendix

\section{\label{sec:GrassmannAlgebra} Grassmann Algebra}

The set of $n$ Grassmann variables, or Grassmann numbers, $g_i$, where $i=1, \dots, n$, are defined to be the generators of the Grassmann algebra, or exterior algebra, of an $n$ dimensional vector space. They compose associatively and distributively, but satisfy the anti-commutation relation
\begin{equation}
    \label{eq:GrassmannAnticommutator}
    \{ g_i, g_j \} = 0 .
\end{equation}

The most general function of Grassmann variables is a multinomial of the form
\begin{equation}
    \label{eq:GrassmannMultinomial}
    F(\bm{g}) = f + \sum\limits_{1 \leq i \leq n} f_i g_i + \sum\limits_{1 \leq i , j \leq n} f_{i, j} g_i g_j + ~\dots ,
\end{equation}
wherein all coefficients are elements of the underlying field, which is taken to be the complex numbers. This definition fails to be unique unless further conditions are specified to accommodate the anti-commutativity. This can be achieved by restricting the summations to follow some prescribed ordering, such as $i < j < \dots$, or by demanding total anti-symmetry in the coefficients themselves so that $f_{i, j, \dots} = \varepsilon_{i, j, \dots} f_{1, 2, \dots}$.

Multinomials in general neither commute nor anti-commute with Grassmann variables or with each other. However, they may always be expressed as $F = F_E + F_O$, where $F_E g_i = g_i F_E$ and contains all the even summands and $F_O g_i = - g_i F_O$ and contains all the odd summands.

Further defined is another set of Grassmann variables $g_i^*$ taken as conjugate companions of the former. The two sets are independent~\cite{Berezin1966, Altland&Simons2010} and anti-commute amidst themselves as well as between each other. They map via conjugation which is an involution that acts accordingly
\begin{equation}
    \label{eq:GrassmannConjugation}
    (g_i g_j \dots g_u g_v)^* = g_v^* g_u^* \dots g_j^* g_i^* , \ (g_i^*)^* = g_i .
\end{equation}

\section{\label{sec:GrassmannCalculus} Grassmann Calculus}

It is possible to extend the notions of calculus to deal with Grassmann quantities. However, their anti-commuting nature makes it is difficult to assign the usual interpretations associated with calculus over the real or complex numbers to such relations. What is obtained is a formal system wherein differentiation and integration act equivalently rather than as inverses of one another.

\subsection{\label{subsec:GrassmannDifferentiation} Grassmann Differentiation}

Grassmann differentiation may act either from the left or from the right, unlike regular differentiation which acts identically in both cases, under these rules
\begin{eqnarray}
\begin{gathered}
    \label{eq:GrassmannDifferentiationRules}
    \overrightarrow{\frac{\partial}{\partial g_i}} g_j = \delta_{i, j} ~, (\overrightarrow{\frac{\partial}{\partial g_i}} g_j F) = (\overrightarrow{\frac{\partial}{\partial g_i}} g_j) F - g_j (\overrightarrow{\frac{\partial}{\partial g_i}} F) , \\
    g_j \overleftarrow{\frac{\partial}{\partial g_i}} = \delta_{i, j} ~, (F g_j \overleftarrow{\frac{\partial}{\partial g_i}}) = F (g_j \overleftarrow{\frac{\partial}{\partial g_i}}) - (F \overleftarrow{\frac{\partial}{\partial g_i}}) g_j , \\
    \overrightarrow{\frac{\partial}{\partial g_i}} 1 = 1 \overleftarrow{\frac{\partial}{\partial g_i}} = 0 .
\end{gathered}
\end{eqnarray}
Informally stated, the anti-commutative property of the Grassmann variables is used to move $g_i$ to either the leftmost or rightmost position in the expression and then it cancels with the derivative operator. Alternatively, the derivative operator itself may be treated as anti-commuting with the Grassmann variables until it is allowed to reach its associated $g_i$ and cancel it.

This definition of Grassmann differentiation implies the following useful results, in which $F_X$ is either even or odd with $\mathrm{sign}(X)$ being $+1$ or $-1$ respectively
\begin{equation}
    \label{eq:LeftDerivativeAntiCommutativity}
    \overrightarrow{\frac{\partial}{\partial g_i}} (\overrightarrow{\frac{\partial}{\partial g_j}} F) = - \overrightarrow{\frac{\partial}{\partial g_j}} (\overrightarrow{\frac{\partial}{\partial g_i}} F) ,
\end{equation}
\begin{equation}
    \label{eq:RightDerivativeAntiCommutativity}
    (F \overleftarrow{\frac{\partial}{\partial g_j}}) \overleftarrow{\frac{\partial}{\partial g_i}} = - (F \overleftarrow{\frac{\partial}{\partial g_i}}) \overleftarrow{\frac{\partial}{\partial g_j}} ,
\end{equation}
\begin{equation}
    \label{eq:LeftDerivativeProduct}
    (\overrightarrow{\frac{\partial}{\partial g_i}} F_X H) = (\overrightarrow{\frac{\partial}{\partial g_i}} F_X) H + \mathrm{sign}(X) F_X (\overrightarrow{\frac{\partial}{\partial g_i}} H) ,
\end{equation}
\begin{equation}
    \label{eq:RightDerivativeProduct}
    (H F_X) \overleftarrow{\frac{\partial}{\partial g_i}} = H (F_X \overleftarrow{\frac{\partial}{\partial g_i}}) + \mathrm{sign}(X) (H \overleftarrow{\frac{\partial}{\partial g_i}}) F_X ,
\end{equation}
\begin{equation}
    \label{eq:DerivativeSwap}
    \overrightarrow{\frac{\partial}{\partial g_i}} F_X = - \mathrm{sign}(X) F_X \overleftarrow{\frac{\partial}{\partial g_i}} .
\end{equation}

Though it might seem arbitrary, as Grassmann variables lack a natural notion of scale, it proves useful to consider the deformation of a Grassmann function $\delta F$ under the transformation $\bm{g} \to \bm{g} +\bm{\delta g}$. The $\delta g_i$ themselves are Grassmann variables, or even more generally any Grassmann odd function. An expression reminiscent of a Taylor series is obtained, where careful attention ought to be paid to ensure the correct ordering of the terms in these relations
\begin{gather}
    \label{eq:LeftTaylor}
    \delta F = \sum\limits_{1 \leq i \leq n} \delta g_i \overrightarrow{\frac{\partial}{\partial g_i}} F + \frac{1}{2!} \sum\limits_{1 \leq i, j \leq n} \delta g_i \delta g_j \overrightarrow{\frac{\partial}{\partial g_j}} \overrightarrow{\frac{\partial}{\partial g_i}} F + ~\dots , \\
    \label{eq:RightTaylor}
    \delta F = \sum\limits_{1 \leq i \leq n} F \overleftarrow{\frac{\partial}{\partial g_i}} \delta g_i + \frac{1}{2!} \sum\limits_{1 \leq i, j \leq n} F \overleftarrow{\frac{\partial}{\partial g_i}} \overleftarrow{\frac{\partial}{\partial g_j}} \delta g_j \delta g_i + ~\dots \ .
\end{gather}

\subsection{\label{subsec:GrassmannIntegration} Grassmann Integration}

Grassmann integration too may act from either direction, though the present focus is on left integration as it ties more closely with the phase space theory. It obeys the following rules
\begin{eqnarray}
\begin{gathered}
    \label{eq:GrassmannIntegrationRules}
    \int dg_i \ g_j = \delta_{i, j} ~, \int dg_i \ 1 = 0 , \\
    \left (\int dg_i \ g_j F \right ) = \left (\int dg_i \ g_j \right ) F - g_j \left (\int dg_i \ F \right ) ,
\end{gathered}
\end{eqnarray}
where the parentheses indicate which terms are being included in the integrand. These are entirely analogous to the rules governing left differentiation in Eqs.~(\ref{eq:GrassmannDifferentiationRules}).
Differentials may be treated as anti-commuting between themselves and the Grassmann variables, and nested integrals are evaluated from the innermost expression
\begin{equation}
    \label{eq:IntegralAntiCommutativity}
    \int dg_i \left (\int dg_j \ F \right ) = - \int dg_j \left (\int dg_i \ F \right ) ,
\end{equation}
maintaining consistency with the anti-commutativity stated in Eq.~(\ref{eq:LeftDerivativeAntiCommutativity}).

Furthermore, as the derivative of a Grassmann function with respect to the Grassmann variable $g_i$ does not contain any instance of said variable, integrating Eq.~(\ref{eq:LeftDerivativeProduct}) and Eq.~(\ref{eq:RightDerivativeProduct}) hence gives
\begin{equation}
    \label{eq:LeftByParts}
    \int dg_i \ (\overrightarrow{\frac{\partial}{\partial g_i}} F_X) H = - \mathrm{sign}(X) \int dg_i \ F_X (\overrightarrow{\frac{\partial}{\partial g_i}} H) ,
\end{equation}
\begin{equation}
    \label{eq:RightByParts}
    \int dg_i \ H (F_X \overleftarrow{\frac{\partial}{\partial g_i}}) = - \mathrm{sign}(X) \int dg_i \ (H \overleftarrow{\frac{\partial}{\partial g_i}}) F_X .
\end{equation}

\onecolumngrid

\section{\label{sec:DifferentialEquations} Phase Space Dynamics}

To show how the differential equations arise, the two components of the Hamiltonian in Eq.~(\ref{eq:HubbardHamiltonian}) are considered in turn and the correspondences in Eqs.~(\ref{eq:P_correspondences}) are applied. The kinetic part produces
\begin{gather*}
    \sum\limits_{i,j, \sigma} \bm{\tau}_{i, j} \hat{a}_{i, \sigma}^\dagger \hat{a}_{j, \sigma} \hat{\rho} \to \sum\limits_{i,j, \sigma} \bm{\tau}_{i, j} (\overrightarrow{\frac{\partial}{\partial g_{i \sigma}}} - \tilde{g}_{i \sigma}) g_{j \sigma} P = \sum\limits_{i,j, \sigma} \bm{\tau}_{i, j} (P g_{j \sigma} \overleftarrow{\frac{\partial}{\partial g_{i \sigma}}} - P \tilde{g}_{i \sigma} g_{j \sigma}) , \\
    \sum\limits_{i,j, \sigma} \bm{\tau}_{i, j} \hat{\rho} \hat{a}_{i, \sigma}^\dagger \hat{a}_{j, \sigma} \to \sum\limits_{i,j, \sigma} \bm{\tau}_{i, j} P \tilde{g}_{i \sigma} (\overleftarrow{\frac{\partial}{\partial \tilde{g}_{j \sigma}}} - g_{j \sigma}) = \sum\limits_{i,j, \sigma} \bm{\tau}_{i, j} (P \tilde{g}_{i \sigma} \overleftarrow{\frac{\partial}{\partial \tilde{g}_{j \sigma}}} - P \tilde{g}_{i \sigma} g_{j \sigma}),
\end{gather*}
where the even nature of the $P$-distribution has been used to apply the derivative swap in Eq.~(\ref{eq:DerivativeSwap}). Hence, taking the sum of these two terms gives
\begin{equation}
    \label{eq:KineticNormalised}
    \sum\limits_{i,j, \sigma} \bm{\tau}_{i, j} (P g_{j \sigma} \overleftarrow{\frac{\partial}{\partial g_{i \sigma}}} + P \tilde{g}_{i \sigma} \overleftarrow{\frac{\partial}{\partial \tilde{g}_{j \sigma}}} -2 P \tilde{g}_{i \sigma} g_{j \sigma}) .
\end{equation}
The interaction potential follows, it leads to some prolonged expressions that are possible to collapse down
\begin{gather*}
    \sum\limits_{i, j, \sigma, \sigma '} \bm{U}_{i \sigma, j \sigma '} \hat{a}_{i, \sigma}^\dagger \hat{a}_{j, \sigma '}^\dagger \hat{a}_{j, \sigma '} \hat{a}_{i, \sigma} \hat{\rho} \to \sum\limits_{i, j, \sigma, \sigma '} \bm{U}_{i \sigma, j \sigma '} (\overrightarrow{\frac{\partial}{\partial g_{i \sigma}}} - \tilde{g}_{i \sigma}) (\overrightarrow{\frac{\partial}{\partial g_{j \sigma '}}} - \tilde{g}_{j \sigma '}) g_{j \sigma '} g_{i \sigma} P , \\
    \sum\limits_{i, j, \sigma, \sigma '} \bm{U}_{i \sigma, j \sigma '} \hat{\rho} \hat{a}_{i, \sigma}^\dagger \hat{a}_{j, \sigma '}^\dagger \hat{a}_{j, \sigma '} \hat{a}_{i, \sigma} \to \sum\limits_{i, j, \sigma, \sigma '} \bm{U}_{i \sigma, j \sigma '} P \tilde{g}_{i \sigma} \tilde{g}_{j \sigma '} (\overleftarrow{\frac{\partial}{\partial \tilde{g}_{j \sigma '}}} - g_{j \sigma '}) (\overleftarrow{\frac{\partial}{\partial \tilde{g}_{i \sigma}}} - g_{i \sigma}) ,
\end{gather*}
considering the cross terms, it is found after re-arranging that
\begin{gather*}
    \bm{U}_{i \sigma, j \sigma '} \tilde{g}_{i \sigma} \overrightarrow{\frac{\partial}{\partial g_{j \sigma '}}} g_{j \sigma '} g_{i \sigma} P = \bm{U}_{i \sigma, j \sigma '} \overrightarrow{\frac{\partial}{\partial g_{j \sigma '}}} \tilde{g}_{i \sigma} g_{i \sigma} g_{j \sigma '} P , \\
    \bm{U}_{i \sigma, j \sigma '} P \tilde{g}_{i \sigma}  \tilde{g}_{j \sigma '} \overleftarrow{\frac{\partial}{\partial \tilde{g}_{j \sigma '}}} g_{i \sigma} = \bm{U}_{i \sigma, j \sigma '} P \tilde{g}_{j \sigma '} \tilde{g}_{i \sigma} g_{i \sigma} \overleftarrow{\frac{\partial}{\partial \tilde{g}_{j \sigma '}}} ,
\end{gather*}
then noting $\bm{U}_{i \sigma, j \sigma '} = \bm{U}_{j \sigma ', i \sigma}$, switching the direction of the derivatives as per Eq.~(\ref{eq:DerivativeSwap}), and combining all the terms
\begin{equation}
    \label{eq:InteractionNormalised}
    \sum\limits_{i, j, \sigma, \sigma '} \bm{U}_{i \sigma, j \sigma '} (P g_{i \sigma} g_{j \sigma '} \overleftarrow{\frac{\partial}{\partial g_{j \sigma '}}} \overleftarrow{\frac{\partial}{\partial g_{i \sigma}}} + P \tilde{g}_{i \sigma} \tilde{g}_{j \sigma '} \overleftarrow{\frac{\partial}{\partial \tilde{g}_{j \sigma '}}} \overleftarrow{\frac{\partial}{\partial \tilde{g}_{i \sigma}}} - 2 P \tilde{g}_{j \sigma '} g_{j \sigma '} g_{i \sigma} \overleftarrow{\frac{\partial}{\partial g_{i \sigma}}} - 2 P \tilde{g}_{i \sigma} \tilde{g}_{j \sigma '} g_{j \sigma '} \overleftarrow{\frac{\partial}{\partial \tilde{g}_{i \sigma}}} + 2 P \tilde{g}_{i \sigma} \tilde{g}_{j \sigma '} g_{j \sigma '} g_{i \sigma}) .
\end{equation}
Finally, bringing Eqs.~(\ref{eq:KineticNormalised}) and~(\ref{eq:InteractionNormalised}) together as prescribed in Eq.~(\ref{eq:GrandCanonicalEnsemble}) gives
\begin{equation}
    \begin{aligned}
    \label{eq:ThermalNormalised}
    \frac{d}{d\beta} P = - \frac{1}{2} &[ - \sum\limits_{i, j, \sigma} \bm{\tau}_{i, j} (P g_{j \sigma} \overleftarrow{\frac{\partial}{\partial g_{i \sigma}}} + P \tilde{g}_{i \sigma} \overleftarrow{\frac{\partial}{\partial \tilde{g}_{j \sigma}}} -2 P \tilde{g}_{i \sigma} g_{j \sigma})
    + \sum\limits_{i, j, \sigma, \sigma '} \bm{U}_{i \sigma, j \sigma '} (P g_{i \sigma} g_{j \sigma '} \overleftarrow{\frac{\partial}{\partial g_{j \sigma '}}} \overleftarrow{\frac{\partial}{\partial g_{i \sigma}}} \\
    & + P \tilde{g}_{i \sigma} \tilde{g}_{j \sigma '} \overleftarrow{\frac{\partial}{\partial \tilde{g}_{j \sigma '}}} \overleftarrow{\frac{\partial}{\partial \tilde{g}_{i \sigma}}} - 2 P \tilde{g}_{j \sigma '} g_{j \sigma '} g_{i \sigma} \overleftarrow{\frac{\partial}{\partial g_{i \sigma}}} - 2 P \tilde{g}_{i \sigma} \tilde{g}_{j \sigma '} g_{j \sigma '} \overleftarrow{\frac{\partial}{\partial \tilde{g}_{i \sigma}}} + 2 P \tilde{g}_{i \sigma} \tilde{g}_{j \sigma '} g_{j \sigma '} g_{i \sigma})] .
    \end{aligned}
\end{equation}

The other differential equation is found by noting that that the Hubbard Hamiltonian in Eq.~(\ref{eq:HubbardHamiltonian}) is number conserving and that the $B$-distribution correspondences in Eqs.~(\ref{eq:B_correspondences}) only change the total Grassmann order by either $+1$ or $-1$. Consequentially, omitting the non-zero order terms in Eq.~(\ref{eq:ThermalNormalised}) then directly gives
\begin{equation}
    \label{eq:ThermalUnNormalised}
    \frac{d}{d\beta} B = - \frac{1}{2} [- \sum\limits_{i, j, \sigma} \bm{\tau}_{i, j} (B g_{j \sigma} \overleftarrow{\frac{\partial}{\partial g_{i \sigma}}} + B \tilde{g}_{i \sigma} \overleftarrow{\frac{\partial}{\partial \tilde{g}_{j \sigma}}}) + \sum\limits_{i, j, \sigma, \sigma '} \bm{U}_{i \sigma, j \sigma '} (B g_{i \sigma} g_{j \sigma '} \overleftarrow{\frac{\partial}{\partial g_{j \sigma '}}} \overleftarrow{\frac{\partial}{\partial g_{i \sigma}}} + B \tilde{g}_{i \sigma} \tilde{g}_{j \sigma '} \overleftarrow{\frac{\partial}{\partial \tilde{g}_{j \sigma '}}} \overleftarrow{\frac{\partial}{\partial \tilde{g}_{i \sigma}}})] .
\end{equation}

\twocolumngrid

\bibliography{Grassmann}

\end{document}